\documentstyle[prl,aps,multicol,epsf]{revtex}

\begin{document}

\author{Mario Castro$^{1}$, Rodolfo Cuerno$^{2}$, Angel 
S\'anchez$^{2}$, and Francisco Dom\'{\i}nguez-Adame$^{1}$}

\address{$^1$GISC, Departamento de F\'{\i}sica de Materiales\\ Facultad de Ciencias
F\'{\i}sicas, Universidad Complutense, E-28040 Madrid, Spain\\ 
$^2$GISC, Departamento de Matem\'aticas\\ Escuela Polit\'ecnica Superior, Universidad Carlos III de
Madrid, E-28911 Legan\'es, Madrid, Spain }

\title{Anomalous scaling in a non local growth model in the Kardar-Parisi-Zhang
universality class}

\maketitle

\begin{abstract}

We study the interface dynamics of a discrete model previously shown [A.\
S\'anchez, M.\ J.\ Bernal, and J.\ M.\ Riveiro, Phys.\ Rev.\ E {\bf 50},
R2427 (1994)] to quantitatively describe electrochemical deposition
experiments.  The model allows for a finite density of biased random
walkers which irreversibly stick onto a substrate.  There is no surface
diffusion.  Extensive numerical simulations indicate that the interface
dynamics is unstable at early times, but asymptotically displays the
scaling of the Kardar-Parisi-Zhang universality class.  During the time
interval in which the surface is unstable, its power spectrum is anomalous;
hence the behaviors at length scales smaller than or comparable with the
system size are described by different roughness exponents.  These results
are expected to apply to a wide range of electrochemical deposition
experiments.

\end{abstract}


\begin{multicols}{2}



The field of non equilibrium dynamics of rough surfaces and interfaces has
undergone an enormous activity during the last decade.  Considerable
experimental and theoretical efforts have been devoted to understanding the
common general features of the growth of surfaces in various seemingly
unrelated phenomena \cite{vicsek,Barabasi}.  By exploiting the existence of
universality and scale invariance akin to those present in equilibrium
critical phenomena, most of the activity has been aimed to organizing non
equilibrium surface growth processes into universality classes.  In this
context, surfaces are described by the {\em global} width (or {\em
roughness}) $W(L,t)$, the rms fluctuations of the height variable $h(x,t)$
giving the location of the surface at position $x$ over a substrate of
lateral size $L$ at time $t$, around its mean value $\bar{h}_L(t)=
\frac{1}{L} \sum_x h(x,t) $:
\begin{equation}
W^2(L,t) = \frac{1}{L} \langle \sum_x (h(x,t) - \bar{h}_L(t))^2 \rangle,
\label{Wtot}
\end{equation}
where angular brackets stand for noise average.

In many cases it is observed that the width initially grows as $W(L,t) \sim
t^{\beta}$ for $t \ll L^z$, saturating at a size dependent value $W(L) \sim
L^{\alpha}$ for $t \gg L^{z}$.  The roughness exponent $\alpha$, the
dynamic exponent $z$ and their ratio $\beta = \alpha/z$ are usually taken
to identify the universality class the growth process considered belongs
to.  Under the assumption of scale invariance (self-affinity), analogous
behavior is expected for the rms fluctuations of the height within a box of
lateral size $l \ll L$, or {\em local} width
\begin{equation}
w^2(l,t) = \frac{1}{l} \langle \sum_x (h(x,t) - \bar{h}_l(t))^2 \rangle ,
\end{equation}
namely
\begin{equation}
w(l,t) \sim \left\{ \begin{array}{lll}
t^{\beta} & {\rm if} & t \ll l^z, \\
l^{\alpha} & {\rm if} & t \gg l^z. \\
\end{array} \right. \label{wlfv}
\end{equation}
The scaling behavior (\ref{wlfv}) holds, {\em e.g}.,\ in the
Kardar-Parisi-Zhang (KPZ) equation \cite{kpz}, which generically describes
surfaces growing irreversibly in the absence of specific conservation laws
\cite{Barabasi}.  However, in recent studies \cite{sch,mbeanom} of surface
growth in the context of Molecular Beam Epitaxy (MBE) and related vapor
deposition techniques, it has been found that many relevant discrete models
and continuum equations do {\em not} obey the local scaling (\ref{wlfv}),
but rather for intermediate times $l^z \ll t \ll L^z$ it is found that
$w(l,t) \sim l^{\alpha_{loc}} \; t^{(\alpha - \alpha_{loc})/z}$, with
$\alpha_{loc} \neq \alpha$; hence the behaviors of the surface at small and
large length scales differ.  This phenomenon has been termed {\em anomalous
scaling}, and is not yet understood at a fundamental level (see
\cite{dasrev,krugrev} for overviews).  In principle, anomalous scaling was
somehow associated with super-roughening ({\em i.e}.,\ the existence of
exponent values $\alpha \geq 1$).  However, recent works have shown that it
can appear as an independent phenomenon \cite{lack} (henceforth referred to
as {\em intrinsic} anomalous scaling) caused by anomalous behavior of the
surface power spectrum \cite{sch,us} (to be defined below).  Up to now, the
only cases in which anomalous scaling has been identified are related to
models with a conserved current along the surface, as those for MBE growth
\cite{dasrev,krugrev}, or with some type of disorder
\cite{lan,lack,us,anomquench,fract}.  We are aware of no previous report on
the occurrence of intrinsic anomalous scaling in models with time dependent
noise and no constraint or conservation law.  Moreover, an obviously
important open issue is to clarify the experimental conditions under which
anomalous scaling can be observable.

In this Rapid Communication we study a surface growth model, called
multiparticle biased diffusion limited aggregation (MBDLA), introduced in
\cite{Angel1} and shown to describe quantitatively electrochemical
deposition (ECD) experiments \cite{Riveiro}.  In particular, it allows to
study systems with a finite density of depositing particles of different
species.  The relaxation rules of MBDLA do not yield a conserved current.
Specifically, consistent with a range of experimental applications
\cite{Riveiro}, there is no surface diffusion.  As a consequence, we will
show below that MBDLA asymptotically displays the scaling behavior of the
KPZ universality class.  Nevertheless, there does exist a time regime in
the evolution of the surface during which the effective scaling is
intrinsically anomalous and described by non universal exponents.  This
time regime is associated with the onset of an instability due to the
Laplacian character of the model.  This feature is analogous to recent
reports \cite{sarmanew} that for a class of model discretized equations in
which there exists a conserved current, anomalous scaling with non
universal exponents is present during a transient regime in which the
surface growth is unstable.  Here we find an analogous phenomenon for a
{\em realistic} model {\em without} conserved currents which belongs to the
KPZ universality class.  Actually, our findings are reminiscent of a
discrete non conserved erosion model \cite{erosion} also displaying an
initial instability followed by asymptotic KPZ scaling.  The crucial
difference is that the model in \cite{erosion} is described by the noisy
Kuramoto-Sivashinsky equation, asymptotically equivalent to KPZ in 1+1
dimensions \cite{krugrev}, and ---as KPZ itself--- {\em free} of anomalous
scaling.


The algorithm that defines MBDLA begins with a number of random walkers
({\em cations}) randomly distributed with concentration $c$ on a
two-dimensional square lattice with periodic boundary conditions.  The
lower side of the lattice is chosen to be the cathode.  The initial
condition evolves in time as follows:  Every time step a walker is chosen
and moved to one of its four neighboring sites with different
probabilities:  $0.5\/$ to move parallel to the cathode (either left or
right); $0.25+p\/$ to move down towards the aggregate (the cathode, at time
$t=0$) and $0.25-p\/$ to move up, away from it.  The parameter $p\/$ is
referred to as the {\em bias} and, as shown in \cite{Angel1}, can be
quantitatively related to the electric current in the physical system;
experimentally, it can be controlled by changing the value of the applied
electric field.  After a destination site has been selected, the particle
moves if that node of the lattice is empty (and we select another particle
if it is not).  Once the particle has been moved, if the new position has
any nearest neighbor site belonging to the aggregate, the
walker$^{\prime}$s present position is added to the aggregate with
probability $s\/$; otherwise it stays there (and is able to move again)
with probability $1-s$.  We term $s\/$ the {\em sticking probability}; it
is related to the chemical activation energy the cation needs to stick to
the aggregate.

Of the two parameters of MBDLA, the bias $p$ and the sticking probability
$s$, only $p\/$ is independent for the experimental description, $s\/$
being simply proportional to it \cite{Angel1}.  This remark
notwithstanding, in this work we will consider both $p\/$ and $s\/$ as
independent parameters, for our goal is to make a general analysis of the
model behavior beyond the experimentally motivated values.  As seen in what
follows, this will help to obtain a coherent, complete picture of MBDLA
properties.  Thus, $s\/$ takes values between $0\/$ and $1$, and $p\/$
ranges from $0\/$ to $0.25$.  Finally, let us note that MBDLA produces in
general a non compact aggregate which can have voids and overhangs.  Large
values of $p$ produce nearly ballistic trayectories, while smaller values
drive the particles in a Brownian motion fashion.  In two limits, the
behavior of MBDLA approaches other relevant models:  ballistic deposition
(for $p = 0.25$), well known to belong to the KPZ universality class and
{\em not} to display anomalous scaling \cite{Barabasi}, and multiparticle
Diffusion Limited Aggregation (MDLA) (for $p = 0$), the paradigmatic model
of unstable Laplacian growth \cite{vicsek}.


To begin the summary of our results, we will describe the phenomenology for
$p=0.05$ and arbitrary $s$.  Physically, $p=0.05$ represents not too small
applied electric fields, surface diffusion being negligible, so that MBDLA
provides a good description of ECD.  We will later discuss how changes in
$p\/$ modify the description that follows.  Simulations were performed in a
$L=300$ by $1500$ lattice, although other values of $L$ were also
considered (see below).  The dynamics of the model is conveniently
characterized by the global width, Eq.\ (\ref{Wtot}). 
\begin{figure}
\narrowtext
\vspace*{-2.05cm}
  \begin{center}
   \setlength{\unitlength}{1cm}
\hspace{-1.2cm}
\begin{picture}(6,6.7)
\epsfxsize=6.3cm
\epsffile{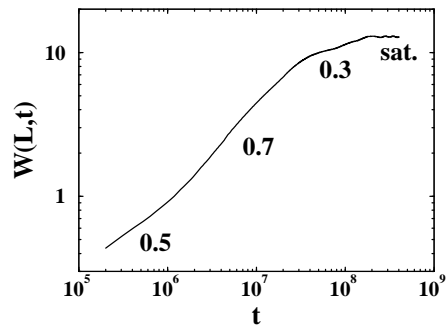}
   \end{picture}    
   \end{center}
 \vspace*{-0.6cm}
\caption{Global width vs time for $p=0.05$, $s=1$, $c=0.1$.  The dynamics
presents an initial transient during which the surface is still
uncorrelated and $\beta \sim \frac{1}{2}$.  Then there is an unstable
transient within which $\beta \sim 0.7 $.  In this region the system
presents {\em intrinsic anomalous scaling} (see text).  Before saturation,
the KPZ nonlinearity governs the dynamics, hence $\beta \sim 0.33$.  The
present plot is an average over $50$ noise realizations.
All magnitudes are measured in arbitrary units.}
\label{regiones}
\end{figure}
We are interested in
the dynamics of the active zone \cite{vicsek}, hence we define the variable
$h(x,t)$ to be the height of the topmost particle belonging to the cluster
for each column.  As mentioned above, for times before saturation due to
finite system size ($t < L^z$), the width is expected to scale as $W(L,t)
\sim t^{\beta}$.  In our case, we find that, along the time evolution,
$\beta\/$ takes up to three different values for large enough systems, see
Fig.\ \ref{regiones}.  At early times, shot noise dominates the growth,
thus $\beta\sim0.5\/$ as in a random deposition process.  As time proceeds,
the Laplacian character of the model shows up, favoring the unstable
development of pillars, which in turn leave deep grooves behind.  This
instability is reflected in the large value of the growth exponent
($\beta\sim0.59-0.73\/$ increasing with sticking probabilities).  As local
slopes of pillars (or grooves) become larger, they seem to trigger further
nonlinear effects, whose consequence is the stabilization of the surface
dynamics.  Thus, for $s\agt 0.5\/$ we measure in this regime
$\beta\sim0.25\/$, a value close to that of the Edwards-Wilkinson
universality class \cite{ew}.  For smaller values of the sticking
probability, $s\alt 0.5$, we measure $\beta\sim0.33\/$, the growth exponent
characteristic of the KPZ universality class.  Actually simulations for
larger system sizes ($L=512$, $1024$) yield $\beta\sim0.33\/$ before
saturation for almost all values of $s$.  Hence the asymptotic behavior of
the model is described by KPZ scaling.  This also shows the role of $s$ as
a noise reduction parameter \cite{kertesz}, large noise reduction ($s\to
0$) tuning the system closer to the asymptotic scaling regime.  The value
of the roughness exponents are $\alpha_{loc}\sim\alpha\sim0.5\/$ in all
cases, showing that there is no anomalous scaling in the asymptotic state.
Measurements of the mean excess velocity $v(m)$ of an interface subject to
an average tilt $m$, imposed through helical boundary conditions, confirm
\cite{Krug} the relevance of the KPZ nonlinearity in the effective
description of our system.  Once established that the long time scaling
behavior of the interface is of the KPZ type, we focus on the unstable
transient regime, characterized by two main features, namely a large
$\beta\/$ value (hence very rapid growth) and {\em intrinsic} anomalous
scaling caused by a non standard form of the power spectrum of the surface,
$S(k,t) = \langle \widehat{h}(k,t) \widehat{h}(-k,t) \rangle$, where
$\widehat{h}(k,t) = L^{-1/2} \sum_x [ h(x,t) - \bar{h}_L(t)] \exp({\rm
i}kx)$. 
\begin{figure}
\vspace*{-2.05cm}
  \begin{center}
   \setlength{\unitlength}{1cm}
\hspace{-1.2cm}
\begin{picture}(6,6.7)
\epsfxsize=6.3cm
\epsffile{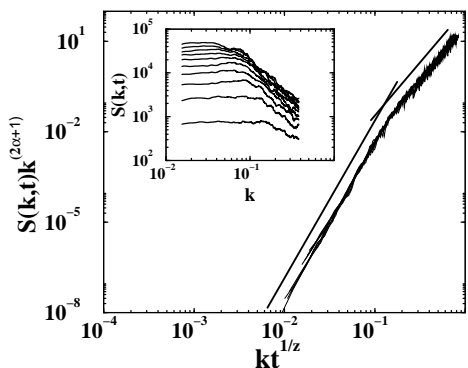}
   \end{picture}    
   \end{center}
 \vspace*{-0.6cm}
\caption{Collapsed power spectrum within the unstable transient using the same 
parameters as in Fig.\ \ref{regiones}. The inset shows the time evolution of the 
power spectrum every $10^6$ time units.  
Thick lines have slopes given by the exponents quoted in Table I.
All magnitudes are measured in arbitrary units.}
\label{power}
\end{figure}
In Fig.\ \ref{power} we plot the colapse of the power 
spectrum $S(k,t)$ within the unstable transient, where the inset shows $S(k,t)$ 
as function of k for every $10^6$ time units.  
$S(k,t)$ displays a behavior consistent with the scaling form
\begin{mathletters}
\label{Sanom}
\begin{equation}
S(k,t)=k^{-(2\alpha+1)}s(kt^{1/z}) , 
\label{Sanom1}
\end{equation}
where
\begin{equation}
s(u)=\left\{\begin{array}{lll}
u^{2\theta} & {\rm if } & u\gg 1,\\
u^{2\alpha+1} & {\rm if } & u\ll 1.\\
\end{array} \right.
\label{Sanom2}
\end{equation}
\end{mathletters}
The exponent $\theta=\alpha-\alpha_{loc}\/$ measures the difference between
the global roughness exponent $\alpha\/$ and the local roughness exponent
$\alpha_{loc}$.  
The exponents are $\alpha\sim2.15$, $\alpha_{loc}\sim0.5$, $\beta\sim0.73$, and
$z\sim2.95$ (all exponent values given are to within an accuracy less than
$5\%$).  
It has been shown in Ref.\ \cite{us} that a structure
factor such as (\ref{Sanom}) implies {\em intrinsic} anomalous scaling for
the local width.  
The slopes of the collapsed curves yield exponents consistent with the ones used to
achieve the best collapse.  Collapses of structure factor data obtained for
different $s\/$ values yield similar qualitative results, as shown in Fig.\
\ref{exps}, and the exponent values summarized in Table I.  All the
exponent values (except notably $\alpha_{loc}$, see also
\cite{dasrev,lack,us}) depend strongly on model parameters, and we consider
them as non universal effective values.
\begin{figure}
\vspace*{-2.05cm}
  \begin{center}
   \setlength{\unitlength}{1cm}
\hspace{-1.2cm}
\begin{picture}(6,6.7)
\epsfxsize=6.3cm
\epsffile{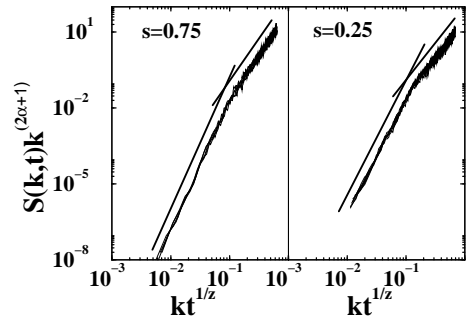}
   \end{picture}    
   \end{center}
 \vspace*{-0.6cm}
\caption{Collapsed power spectra within the unstable transient using the
exponent values in Table I, with $s=0.75$ and $s=0.25$.  Parameters other than $s$ are as in Fig.\
\ref{regiones}. 
Thick lines have slopes given by the exponents quoted in Table I.
All magnitudes are measured in arbitrary units.}
\label{exps}
\end{figure}

To provide a general description of our model, we have studied {\em bias}
values different from $p=0.05$.  For example, for $p=0.005\/$ and large
values of the sticking probability $s\/$, we have observed DLA like
patterns as expected \cite{Angel1}.  Consequently the unstable regime lasts
longer than in the cases studied above.  The same is observed for larger
values of $s$ at low $p$ values.  Moreover, decreasing (increasing) the ion
concentration $c$ increases (decreases) the duration of the unstable
growth, in accordance with the fact that as $c\to 0$ pure DLA is recovered.
Hence we can tune the two parameters and the concentration to reproduce
different experimental behaviors, because changes in the parameters alter
the relative duration of the different scaling regions in the dynamics of
the system.  However the asymptotic KPZ scaling still holds at long enough
times.


In summary, we have shown the existence of intrinsic anomalous scaling for
a realistic model, MBDLA, without conserved currents and hence in the KPZ
universality class.  Anomalous scaling appears within a time region of
variable duration (depending on parameter values) in which the surface
evolves in an unstable fashion.  This type of scaling is due to a
non-standard form of the power spectrum and features non-universal
(parameter dependent) values of the effective exponents.  We interpret
anomalous scaling to be associated with the unstable structures developing
in the model dynamics.  Thus, for large length scales we measure
$\alpha>1\/$, associated with the steep slopes of pillars.  On the other
hand, when probing short length scales we obtain $\alpha_{loc}\sim 0.5\/$
(see Table I), a value numerically close both to the KPZ exponent and to
the exponent resulting from the convolution of the Heaviside function with
the interface \cite{Kah} (through our definition of the height variable
$h(x,t)$).  At any rate, intrinsic anomalous scaling seems to be strongly
associated in our case with the nonlocality of the evolution rules of the
model which produces the unstable structures of the interface.

To date, anomalous scaling has been mostly observed in theoretical models,
hence the interest of our results:  as MBDLA is a correct description of
ECD \cite{Riveiro}, our findings could be observed in that kind of
experiments.  Interestingly, MBDLA displays instabilities similar to those
reported in various ECD experiments \cite{rubio}, reinforcing the
connection of the model to actual physical systems.  Moreover, anomalous
scaling may explain the wide spread in the $\alpha$ values
($\alpha\sim0.5-0.8\/$) reported in various ECD experiments \cite{rubio}.
As shown in \cite{lack}, in the presence of anomalous scaling, the fact
that {\em all} length scales saturate at the same {\em global} saturation
time $t_{sat} \sim L^z$ poses additional difficulties to the evaluation of
exponents by means of the local width.  Specifically, if not performed at
global saturation, the roughness exponent thus obtained is an effective
value different from both, $\alpha\/$ and $\alpha_{loc}$.  Research on
extensions of the model to more experimental parameters is under way, the
influence of surface diffusion being the center of our future work.

\acknowledgments

We are indebted to Enrique Diez for his participation in the early stages
of this work.  We thank Miguel Angel Rodr\'{\i}guez, Miguel Angel Rubio,
Javier Buceta and Ricardo Brito for helpful comments.  This work has been
partially supported by CICyT (Spain) under Project No.  MAT95-0325.

\end{multicols}
\end{document}